\documentclass[doublecol]{epl2} 
\usepackage{stickstootext}
\usepackage[stix2]{newtxmath}
\usepackage[letterspace=12]{microtype} 

\usepackage[pdftex,colorlinks=true,citecolor=blue,urlcolor=blue,linkcolor=black]{hyperref}

\title{Quantum Borderlines}
\subtitle{Fluctuation Energies in Ultracold Bose Gases} 
\shorttitle{Quantum fluctuations of massive Bosons} 

\author{Carsten Henkel}

\institute{%
University of Potsdam, Institute of Physics and Astronomy, Karl-Liebknecht-Str. 24/25, 14476 Potsdam, Germany
}

\abstract{
Ultracold Bose gases are many-body systems with well-defined particle
interactions that may serve as models for interacting quantum fields.
The impact of virtual excitations is studied in the spatial transition zone
created by a soft confinement potential that separates a degenerate
ideal gas from a dense quasi-condensate.
We compute within Bogoliubov theory the contribution of quasi-particles 
to the surface energy at zero temperature.
}


\newcommand{\reftoarxiv}{Appendix}

\usepackage{comment}

\begin{document}

\maketitle

\section{Introduction}

Since the inception of the kinetic theory of heat \cite{Maxwell_Heat},
microscopic fluctuations like the velocity of gas particles 
are recognised in averaged quantities like heat or pressure.
In the quantum theory, fluctuations or often called indeterminacies
persist even at zero temperature (in the ground state) and can be 
quantified by zero-point energies or by the finite spread of typical
observables (e.g., second moments).
Dispersion forces of the van der Waals and Casimir type can be understood
in this way as rectified underlying microscopic fluctuations
\cite{Eisenschitz30}.

In this paper, these ideas are exemplified within a simple non-relativistic 
quantum field theory for an ultracold gas of massive Bosons
\cite{Lewenstein96b, Andersen04}.
Casimir energies for samples trapped in simple geometries have been
discussed in earlier work
\cite{Biswas_2007b, Edery_2006b, Fuchs_2007, Martin_2006,  Gambassi_2006, Recati_2005, Roberts05, Reichert_2019a} and follow the same scaling laws as for scalar fields or
the electromagnetic field, provided the dispersion relation for
elementary excitations is linear in the relevant range.
More general dispersions have been studied by Bachmann and Kempf
\cite{Bachmann_2008} who constructed the Casimir energy for a 
one-dimensional scalar field as a series based on a 
polynomial dispersion relation $\omega(k)$. 
Curiously enough, there is no Casimir interaction for quadratic dispersions 
like for free non-relativistic particles.
For the well-known Bogoliubov dispersion of the Bose gas
[see Eq.\,(\ref{eq:surface-energy-3D-BEC}) below],
an explicit integral representation of the Casimir energy has been given 
\cite{Schiefele_2009} whose scaling deviates for plates at distances
smaller than the healing length $\xi = \hbar/(2mc)$ where $m$ is the 
Boson mass and $c$ the long-wavelength speed of sound.

We study here a simpler, but related quantity, namely the surface energy
of a Bose gas \cite{Gaudin_1971, Graham01, Reichert_2019b}.
As in the case of {droplets formed by} two non-miscible species
\cite{Schaeybroeck08, Boudjemaa21}, the surface energy translates the distortion
of the collective particle wave functions in the interface region.
It has contributions from both the highly occupied ground state, the
so-called condensate mode, and from its elementary excitations. 
It is interesting that the latter even contribute at zero temperature,
similar to what is called depletion where virtual collisions promote
condensate particles to higher orbitals \cite{PitaevskiiStringari, Popov77}.
In this example, the surface energy due to quantum fluctuations 
arises via interactions among the particles \cite{Jaffe05} --
these are in particular responsible for the dispersion relation
changing from quadratic to linear at large wavelengths.
Similar physics appears also in fermionic systems where a part of 
the surface energy of a metal can be traced to collective 
charge fluctuations
(bulk and surface plasmons) \cite{Barton79, MorgensternHoring85}.

The paper starts with a ``sum over mode'' expression for a 
planar surface and then focuses on the border of a one-dimensional
system. 
Its collective modes are calculated numerically for three generic
potentials that define a more or less soft confinement.
The deviations of these modes from a homogeneous system
define the surface energy that we compute
in the ground state (zero temperature), using an expansion in the
number of particles that do not condense in the lowest-energy mode
(i.e., a mean-field theory and its fluctuations up to second or third order
\cite{Mora03, Henkel17b}).

\section{Surface energy in 3D}

In earlier work \cite{Schiefele_2009}, we studied the Casimir geometry
(two plane surfaces) for an ultracold Bose gas and focussed on the
distance-dependent part of the ground state energy. 
This paper also
quoted the surface energy for Dirichlet boundary conditions,
given in terms of
the dispersion relation $\omega(k)$ of elementary excitations 
(also known as Bogoliubov quasi-particles)
\begin{equation}
{\cal E}_D = - \frac{\hbar}{4} \int\!\frac{ {\rm d}^{2}k }{ (2\pi)^{2} }
\omega(k)
\,,\quad
\omega(k)
= c k \sqrt{1 + k^2 \xi^2}
\label{eq:surface-energy-3D-BEC}
\end{equation}
where 
${\bf k}$ is the wavenumber parallel to the surface,
$c = \hbar/(2m\xi)$ is the speed of sound, 
and the healing length $\xi$ relates to the chemical potential,
$\hbar^2/(2m\xi^2) = 2\mu$.
(For Neumann boundary conditions, the surface energy has the opposite sign.)
The integral~(\ref{eq:surface-energy-3D-BEC}) can be renormalised
by subtracting counter terms from the UV limit
\begin{equation}
\omega(k) = \frac{\hbar}{2m} \left( k^2 + \frac{ 1 }{ 2 \xi^2 } - \frac{ 1 }{ 8 k^2 \xi^4 } + \ldots \right)
\label{eq:}
\end{equation}
(see Refs.\cite{AlKhawaja02b, Andersen04, Graham01} for more details).
When the last term is subtracted (or not), the divergence is reduced to
a logarithmic one in the IR (or UV).
With the substitution $k = (1/\xi) \sinh t$, we compute (without the last counter
term)
\vspace{-2ex}
%
\begin{eqnarray}
{\cal E}_3 &=& 
- \frac{\mu}{8\pi \xi^2}
\int\limits_0^\infty\!{\rm d}t\,
\sinh 2t \left(
\sinh t \cosh t - \sinh^2 t - {\textstyle\frac12}
\right)
\nonumber\\
&=&
\frac{\mu}{32\pi \xi^2}
\left( 
\mathop{\rm arsinh} (k_{\rm UV} \xi)
- {\textstyle\frac{1}{4}} \right)
\label{eq:}
\end{eqnarray}
This model is not satisfactory, however, because on physical grounds,
the Dirichlet boundary condition entails more than simply removing 
[as in Eq.\,(\ref{eq:surface-energy-3D-BEC})]
one half mode at zero momentum perpendicular to the plates.
Also the condensate mode function should vanish at the boundary which
leads (in the perpendicular direction) to a Jacobi 
elliptic function $\mathop{\rm sn}$ \cite{Carr00a}.
We also anticipate that the ground state energy of a many-body system 
like the Bose gas has a structure that differs from a quantised
relativistic field.
It vanishes, by construction, for zero particles and shows, at finite
chemical potential, particle and energy fluctuations in the equilibrium state
that minimises the grand-canonical potential.
A special feature introduced by particle interactions in a Bose gas
are ``coherence factors'' or ``Bogoliubov amplitudes'' 
[see Eq.\,(\ref{eq:Ansatz-for-psi}) below] that convey the
particle content of the elementary excitations: at long wavelengths,
these are mainly phase excitations of the condensate, and change to
single-particle character at short wavelengths \cite{PitaevskiiStringari}.
The ground state energy is indeed proportional to the ``hole density''.
Its spatial dependence near the border of the system will be studied now.

\begin{figure}[tbp]
   \centerline{%
   \hspace*{-01mm}
   \includegraphics[width=0.25\textwidth]{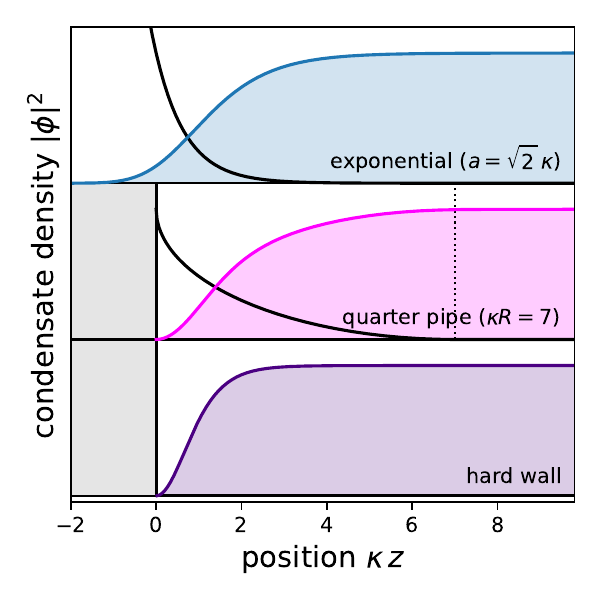} 
   \hspace*{-02mm}
   \includegraphics[width=0.255\textwidth]{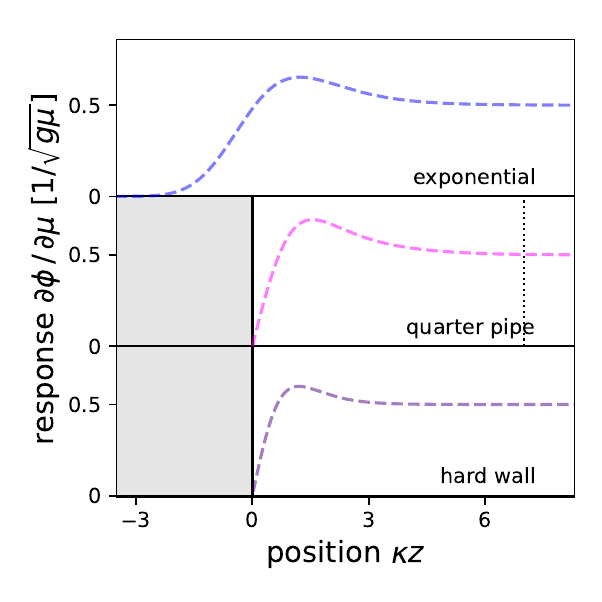} 
   \hspace*{-01mm}%
   }
   \vspace*{-03ex}
   \caption[]{(left) Illustration of condensate density for three model potentials.
   (right) Response of condensate mode to a change in the chemical potential.
   Position scaled to healing length $1/\kappa = \hbar/\sqrt{m\mu} = 2\xi$. 
   Exponential potential (top left, black solid line):
   $V(z) = \mu\,{\rm e}^{-a z}$.
   ``Quarter pipe'': $V(z) = \mu - \mu \sqrt{1 - (1 - z/R)^2}$ for $z \le R$.
   Hard wall: $V(z) = 0$.
   In the last two cases, the boundary condition $\varphi(0) = 0$ is applied.}
   \label{fig:condensates}
\end{figure}

\section{Fluctuating Bose field theory}

We consider a one-dimensional Bose gas in a half-open trap described 
by the potential $V(z)$. 
Three model potential are sketched in Fig.\:\ref{fig:condensates}(left):
the turning point in the classical sense is located at $z = 0$,
but the actual boundary condition depends on the potential (soft or
hard wall).

The field theory for the Bose (quasi)-condensate and its excitations can be built on the
grand-canonical Hamiltonian density
\cite{Hugenholtz_1959, Griffin96, Oehberg97, Fedichev98, Mora03}
\begin{equation}
{\cal H} = - \frac{\hbar^2}{4m} \left(
 \Psi^\dag \frac{ {\rm d}^2\Psi }{ {\rm d}z^2 }
+ \mbox{h.c.} 
\right)
+ (V - \mu) \, \Psi^\dag \Psi
+ \frac{ g }{ 2 } \Psi^{\dag 2} \Psi^2
\label{eq:}
\end{equation}
This form is locally Hermitean and has been put in normal order 
with respect to the field operator $\Psi$, 
$\mu$ is the chemical potential and $g > 0$ the strength of pair
interactions for two Bosons in the same place (contact potential)
\cite{PitaevskiiStringari}. (In the one-dimensional setting, no
pseudopotential is needed for UV convergence.)
The parameter $g$ depends on the confinement in the two other 
directions and the atomic species \cite{Olshanii98};
a typical range is $g \sim 0.3 \ldots 3$\,nK\,µm.

One assumes that the bulk of the particles occupy the condensate mode 
$\varphi$ which is found by minimising ${\cal H}$.
This leads to the nonlinear Schrödinger (Gross-Pitaevskii) equation
\begin{equation}
- \frac{\hbar^2}{2m} 
 \frac{ {\rm d}^2\varphi }{ {\rm d}z^2 }
+ V(z) \, \varphi
+ g |\varphi(z)|^2 \, \varphi
= \mu \, \varphi
\label{eq:GPe}
\end{equation}
Examples are shown in Fig.\:\ref{fig:condensates}, computed numerically,
except for the hard-wall potential which has the analytical solution
$\varphi(z) = (\mu/g)^{1/2} \tanh \kappa z$
\cite{PitaevskiiStringari}.
Assuming that fluctuations around the mean-field condensate are small
enough, one expands for $\Psi = \varphi + \hat{\psi}$ to second
order in the operator $\hat{\psi}$, giving
%
%
%
\begin{eqnarray}
{\cal H}^{(2)} &=& - \frac{\hbar^2}{4m} \left(
  \hat\psi^\dag \nabla^2 \hat\psi
+ \mbox{h.c.} 
\right)
+ (V_2 - \mu) \, \psi^\dag \psi
\nonumber\\
&& + \frac{ g }{ 2 } \left(
  \varphi^{*2} \hat{\psi}^2
+ \mbox{h.c.} 
\right)
\label{eq:Omega-2nd-order}
\end{eqnarray}
with the Hartree-Fock potential $V_2 = V + 2 g |\varphi|^2$.
The last line is a parametric interaction with the condensate 
that drives (or squeezes) fluctuations \cite{Navez98}.


Within this one-dimensional model, the elementary excitations 
can be enumerated by a single frequency parameter $\omega$.
The Boboliubov Ansatz for a continuum of frequencies (as appropriate
in a half-open potential)
\begin{equation}
\hat\psi(z) = \int\!\frac{{\rm d}\omega}{\sqrt{\pi}} \left( u_\omega(z) \, b(\omega) + v_\omega(z) \, b^\dag(\omega) \right)
\label{eq:Ansatz-for-psi}
\end{equation}
contains Bosonic operators with 
$\left[ b(\omega), \, b^\dag(\omega') \right] = \delta(\omega - \omega')$.
This form diagonalises the Hamiltonian~(\ref{eq:Omega-2nd-order})
[see \reftoarxiv~A.1]
provided the “particle” and “hole” wave functions solve the system
(for simplicity, the parameter $\omega$ has been suppressed here) 
\begin{equation}
\begin{aligned}
-\frac{\hbar^2}{2m} \frac{{\rm d}^2 u}{ {\rm d}z^2 } + (V_2 - \mu) \, u + g \varphi^2 \, v^* &=& \hspace*{-1.2ex} \hphantom{-} \hbar\omega \, u
\\
-\frac{\hbar^2}{2m} \frac{{\rm d}^2 v}{ {\rm d}z^2 } + (V_2 - \mu) \, v + g \varphi^2 \, u^* &=& \hspace*{-1.2ex} -\hbar\omega \, v
\end{aligned}
\label{eq:BdG}
\end{equation}
The standard bilinear form used to normalise the continuum modes is
\cite{PitaevskiiStringari, Diallo15a}
\begin{equation}
\int\!\frac{{\rm d}z}{\pi} \left( u^*_\omega u^{}_{\omega'} 
- v^*_\omega v^{}_{\omega'} \right) = \delta(\omega - \omega')
\label{eq:uv-orthogonal}
\end{equation}
When the Ansatz~(\ref{eq:Ansatz-for-psi}) is inserted into Eq.\,(\ref{eq:Omega-2nd-order}) and the Bogoliubov equations are used, the spatial integration yields
(see \reftoarxiv~A.1, primed quantities correspond to the frequency $\omega'$)
\begin{eqnarray}
H^{(2)} &=& \int\!{\rm d}z\, \frac{{\rm d}\omega\, {\rm d}\omega'}{\pi} 
\frac{\hbar\omega}{2} 
\left[
\left( u^{\prime*} \, b^{\prime\dag} + v^{\prime*} \, b' \right)
\left( u \, b - v \, b^\dag \right)
\right.
\nonumber\\
&& \hphantom{{\rm d}z\, {\rm d}E\, {\rm d}E' \frac{\hbar\omega}{2}}
\left. + 
\left( u^* \, b^\dag - v^* \, b \right)
\left( u' \, b' + v' \, b^{\prime\dag} \right)
\right]
\nonumber\\
&=&
\int\!{\rm d}\omega \, \hbar\omega \, b^\dag(\omega) \, b(\omega)
+ {\cal E}^{(2)}
\label{eq:expand-H2-step-1}
\end{eqnarray}
This is the desired canonical form with the quasi particle number
$b^\dag b$ (per frequency) that vanishes in the ground state. 
The ground state energy ${\cal E}^{(2)}$ arises from putting
the terms $-v^* v'\, b b^{\prime\dag}$ into normal order. 
This gives finally
\begin{equation}
{\cal E}^{(0+2)} = \int\!{\rm d}z \left[
- \frac{ g }{ 2 } |\varphi|^4 - 
\int\!\frac{{\rm d}\omega}{\pi} \hbar\omega \, |v|^2
\right]
\label{eq:ground-state-energy-0+2}
\end{equation}
The first term is the contribution of the condensate $\varphi$
to the energy (zeroth order in fluctuations), simplified by 
using Eq.\,(\ref{eq:GPe}).
--
In the following, we analyse the spatial shape of these contributions
with a focus on the border region around $z = 0$.

\begin{figure}[tbp]
   \centering
   \includegraphics[height=0.25\textwidth]{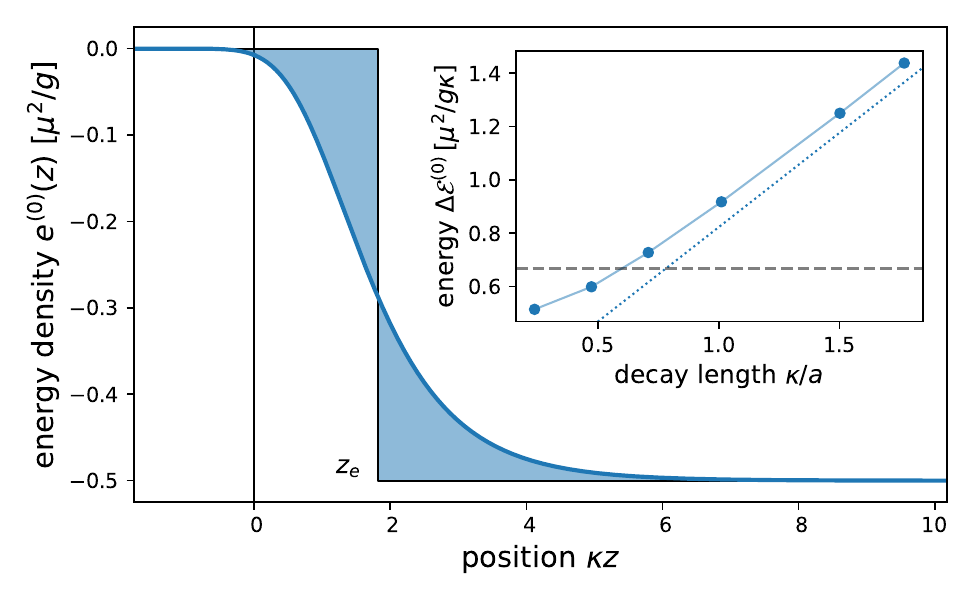}
   \vspace*{-2ex}
   \caption[]{One-dimensional Bose gas near a soft wall set by an
   exponential barrier at $z = 0$.
   Ground state energy density due to the condensate mode
   function (main plot, length parameter $1/a = 1/\kappa$). The position
   $z_e = 2 g \Delta{\cal E}^{(0)} / \mu^2$ marks the fictitious termination
   of a homogeneous system that would give the same ground state energy
   (the two blue shaded areas compensate).
   The inset shows the surface tension $\Delta{\cal E}^{(0)}$ as a function of $1/a$.
   In dashed, the result~(\ref{eq:surface-tension-hard-wall})
   for the hard wall. The dotted line illustrates a linear scaling
   with $1/a$.
   }
   \label{fig:surface-tension-0}
\end{figure}

%
\section{Energy from condensate mode}

The condensate contribution ${\cal E}^{(0)}$
is illustrated in Fig.\:\ref{fig:surface-tension-0}
for the exponential potential. 
Far away from the border, the potential $V(z)$ vanishes, 
and the condensate density $g |\varphi(z)|^2 \to \mu$.
We identify a surface contribution $\Delta {\cal E}^{(0)}$
to the energy in Eq.\,(\ref{eq:ground-state-energy-0+2})
by extracting the part extensive in the system length $L$
and proportional to the bulk energy density 
\begin{equation}
{\cal E}^{(0)}(L) = - \frac{ L \mu^2 }{ 2 g } + \Delta{\cal E}^{(0)}
\label{eq:energy-0-extensive-plus-border}
\end{equation}
An explicit result is found for the hard wall 
\begin{equation}
\Delta{\cal E}^{(0)} = \frac{\mu^2}{2g} \int\limits_0^L\!{\rm d}z
\left( 1 - \tanh^4 \kappa z \right) 
= \frac{2 \mu^2 }{ 3 g \kappa }
\label{eq:surface-tension-hard-wall}
\end{equation}
taking the macroscopic limit $\kappa L \to \infty$. 
This agrees with the leading order obtained within the Lieb-Liniger model 
for weak interactions \cite{Reichert_2019b}.
If this surface energy is compared to the chemical potential (the
energy needed to add one particle to the system), it corresponds
%
to a particle number $\Delta N^{(0)} = 2 \mu / (3 g \kappa)$,
equal to a sample of length $\frac{4}{3}\xi$ at the bulk density.
Typical values depend on the trap geometry and the 
isotope, but a generic range is $\Delta N^{(0)} \sim 8\ldots 100$.
The dependence on the exponential potential
is illustrated in Fig.\:\ref{fig:surface-tension-0}(inset):
the surface energy increases as the potential gets softer,
it roughly scales with the bulk energy density times 
the length parameter $1/a$ (dotted line).
%

\begin{figure}[tbph]
   \centering
   \includegraphics[height=0.4\textwidth]{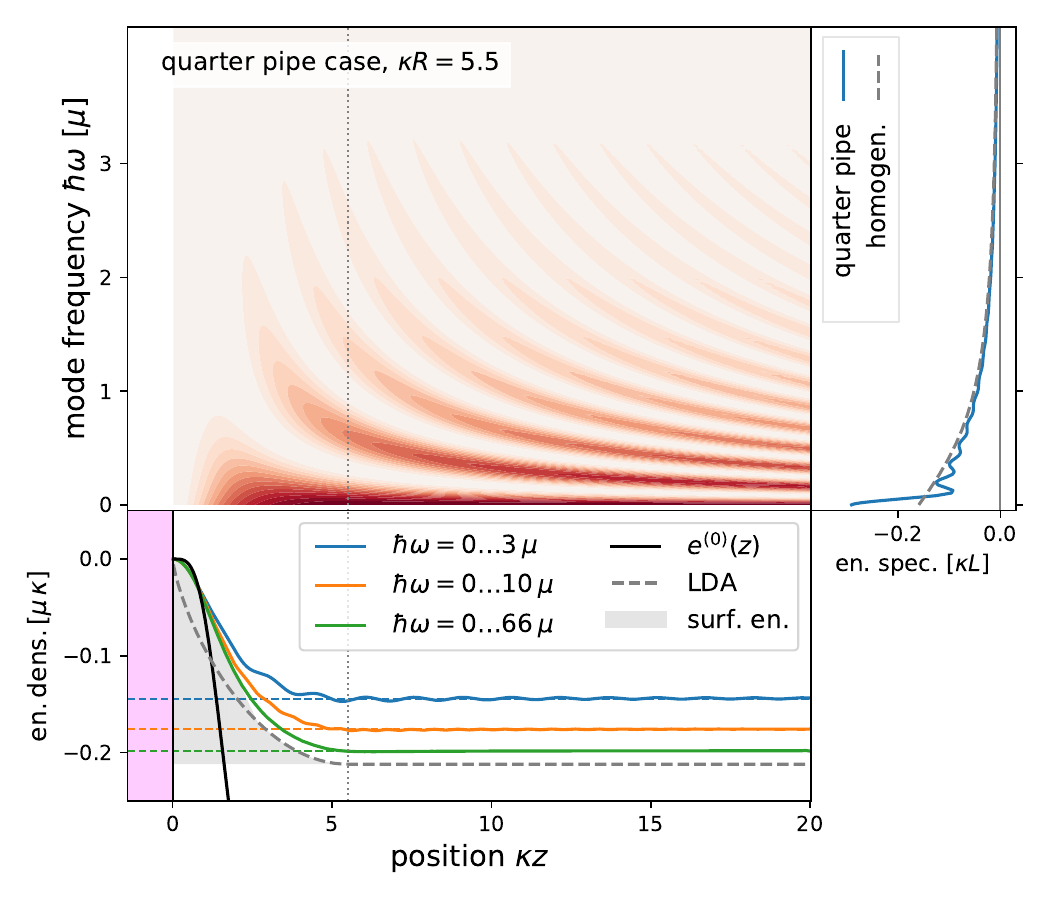} 
   \vspace*{-2ex}
   \caption[]{
   Spectral representation of the ground-state (zero temperature)
   energy density $-\hbar\omega \, |v_\omega(z)|^2$
   in the border region of an inhomogeneous Bose gas for the quarter-pipe
   potential.
   The bottom and right panels give the marginal integrated form
   of the energy density along the energy or position axis, 
   as indicated in the legends.
   Solid black line in bottom plot: contribution from condensate,
   $e^{(0)}(z) = - (g/2) |\varphi(z)|^4$ 
   [Eq.\,(\ref{eq:ground-state-energy-0+2})].
   Shaded area: estimated surface energy 
   [Eq.\,(\ref{eq:surface-energy-quarter-pipe})].
   The upper limits for the spatial and frequency integrations are
   $z \le 20/\kappa, \hbar\omega \le 3, 10, 66\,\mu$ as indicated.
   The results for a locally homogeneous system [where mode functions 
   are spatially averaged, see 
   Eq.\,(\ref{eq:homogeneous-v2})] are plotted in dashed. 
   }   
   \label{fig:energy-density-quarter-pipe}
\end{figure}

\section{Energy from virtual quasi-particles}

Returning to the second-order fluctuation part in
Eq.\,(\ref{eq:ground-state-energy-0+2}), we plot the integrand 
$-\hbar\omega \,|v_\omega(z)|^2$ as a function of position $z$ and mode frequency
$\omega$ in Fig.\:\ref{fig:energy-density-quarter-pipe}. The “quarter-pipe”
potential (central panel of Fig.\:\ref{fig:condensates}) is taken in this 
example.
The mode functions are evaluated numerically using the scheme developed
in Ref.\cite{Diallo15a}.
We note that the infrared limit is regular because of the factor $\omega$.
In the ultraviolet, the integrand smoothly decays because
the ratio $v / u$ gets small 
as the kinetic energy
in Eqs.\,(\ref{eq:BdG}) becomes dominant relative to the $uv$ cross coupling.
(This is similar to the ``upper'' and ``lower'' components of a Dirac
spinor in the non-relativistic limit \cite{Messiah_vol2}.)

The oscillatory behaviour of the wave functions $v_{\omega}(z)$ can be
captured at large distance from the border by \cite{Muryshev99, Kovrizhin01a,  Dziarmaga04, Walczak11}
\begin{equation}
V(z) \to 0: \quad 
v_\omega(z) \to - \frac{ \sin \theta/2 }{ \sqrt{c\,k\xi} } \sin(k z + \delta)
\label{eq:uv-normalisation-flat-bottom}
\end{equation}
with $\cot\theta = \hbar\omega/\mu$ and 
and the Bogoliubov wave number
\begin{equation}
k = \frac{ \omega } { c } \sqrt{ \frac{ 2 \mu }{ \mu + C } }
\label{eq:Bogoliubov-dispersion}
\end{equation}
using $C^2 = \mu^2 + (\hbar\omega)^2$.
The phase shift $\delta = \delta(\omega)$ is studied in a separate paper
\cite{phase-shift-I}.
In this potential, it starts at $\pi/2$ at low frequencies
and slowly drops to zero for $\hbar\omega \gg \mu$, 
qualitatively similar to the hard-wall case.

The panels on the bottom and to the right display the integrals over
energy and over position, respectively.
The spatial integral (right panel in Fig.\:\ref{fig:energy-density-quarter-pipe}) 
scales, in the leading order, with the system length $L$.
This corresponds, as for the condensate contribution
in Eq.\,(\ref{eq:energy-0-extensive-plus-border}),
to the ground state energy density in a homogeneous system. 
The energy integration can be performed explicitly
for $\kappa z \gg 1$
when the hole function oscillations are averaged out
(denoted by the overline)
\begin{equation}
\hbar\omega \, \overline{v^2(z; \omega) }
\approx 
\frac{ \hbar\omega \sin^2 \theta }{ 4 c\, k \xi (1 + \cos\theta)} 
%
= \hbar \kappa \frac{ \sqrt{2 \mu^3 (\mu + C)} }{ 4 C (\hbar\omega + C) }
\label{eq:homogeneous-v2}
\end{equation}
The substitution $\hbar\omega = \mu \sinh 2 t$ reduces the energy density
in second order $e^{(2)}(z)$ to the elementary
\cite{Popov77, Mora03}
\begin{eqnarray}
e_{\rm hom}^{(2)}(\leq \omega) & = 
- \displaystyle
\frac{\mu \kappa}{\pi}
\int_0^t\!{\rm d}t \, \cosh(t) \, {\rm e}^{-2t}
\nonumber\\
&=
- \displaystyle
\frac{ \mu \kappa }{ 2\pi } 
\left( 
	{\textstyle\frac43}
	- {\rm e}^{-t} - {\textstyle\frac{ 1 }{ 3 }} {\rm e}^{-3t} 
\right)
\label{eq:energy-density-3}
\end{eqnarray}
Since ${\rm e}^{-t} \sim \omega^{-1/2}$, the convergence in the UV
is relatively slow.
In the bottom panel, this expression is plotted in dashed and agrees
well with the numerical integrations up to some finite energy,
provided the distance from the border is large enough.
The oscillatory features may suggest Friedel oscillations familiar from
the electron density at a metal surface \cite{Lang69, Kenner72}.
They are actually artefacts of the sharp upper limit in the numerical
data because the integrand is smooth and no oscillatory features are 
to be expected when all frequencies are included.
A relatively good agreement with the full energy integral is obtained
by evaluating Eq.\,(\ref{eq:energy-density-3}) 
in the local density approximation, i.e., inserting 
$\mu(z) = \mu - V(z)$ (gray dashed line in bottom panel).

The contribution of elementary excitations to the surface energy of
the Bose gas can be defined by a construction similar to 
Fig.\:\ref{fig:surface-tension-0}: we subtract the bulk limit and
integrate over the “hole” in the region $z \approx 0 \ldots 10/\kappa$
(light gray shaded area in lower panel of 
Fig.\:\ref{fig:energy-density-quarter-pipe}).
The result is approximately
\begin{equation}
\Delta {\cal E}^{(2)} \approx 0.41\,\mu 
\quad \mbox{(quarter pipe)}
\label{eq:surface-energy-quarter-pipe}
\end{equation}
equivalent to roughly one half atom per boundary. 
Note that compared to the mean-field 
contribution~(\ref{eq:surface-tension-hard-wall}),
this term is smaller by a factor of the order 
$\kappa / n = \sqrt{\gamma}$ where $\gamma$ is the Lieb-Liniger parameter
for the contact interaction \cite{Reichert_2019b, BouchouleChapter11}.

\section{Condensate phase and particle number}

So far, we have taken into account quantum fluctuations of quasi-particles
propagating in the condensate background field.
But in the condensate itself, the particle number fluctuates. 
Both the contact to a particle reservoir (grand-canonical ensemble) 
and the scattering into elementary excitations (quasi-particles) play a role.
It has become customary to write the fluctuation operator on top of
the c-number-valued condensate mode function $\varphi$ in the form
\cite{Lewenstein96a, Gardiner97b, Castin98}
\begin{equation}
\hat{\psi}_c = - {\rm i} \hat{Q} \varphi + \hat{P} \varphi_a
\label{eq:psi-zero-mode}
\end{equation}
Here, the operators $\hat{Q}$, $\hat{P}$ represent phase and number
fluctuations and commute to $[\hat{Q}, \hat{P}] = {\rm i}$. 
The so-called adjoint mode function $\varphi_a$ projects out the
condensate mode according to
\begin{equation}
2 \int^L\!{\rm d}z\, \varphi_a \varphi = 1
\label{eq:phi-a-projector}
\end{equation}
and is orthogonal to the Bogoliubov mode functions for $\omega > 0$,
$\int\!{\rm d}z\, \varphi_a (u_\omega - v_\omega) = 0$
\cite{Dziarmaga04, Castin98}.
To regularise the integrals in the following, 
we have again introduced an upper limit $L$ in the bulk region, 
far from the border.

By taking the $\mu$ derivative of the Gross-Pitaevskii equation, one
finds the explicit solution
\cite{Castin98}
\begin{equation}
\varphi_a = \mu'_0
\frac{ \partial \varphi }{ \partial \mu }
\,,
\qquad
\mu'_0 = \frac{ \partial \mu }{ \partial N } 
\label{eq:phi-a-mu-derivative}
\end{equation}
where $N$ is the number of condensed particles, i.e., the integral of
$|\varphi|^2$, of which 
Eq.\,(\ref{eq:phi-a-projector}) simply is the $N$-derivative.
The bulk value $\varphi \to (\mu/g)^{1/2}$ 
in a flat-bottom potential yields
the constant product
$2 \varphi_a \varphi \to \mu'_0 / g$. In the leading order,
Eq.\,(\ref{eq:phi-a-projector})
thus implies the scaling $\mu'_0 L / g \approx 1$. 
The behaviour of the condensate density near the border is
captured by a correction length,
\begin{equation}
\mu'_0 = \frac{ g }{ L - \Delta L }
\label{eq:}
\end{equation}
Typical values for $\Delta L$, indeed of the order of the healing
length $1/\kappa$, are given in Table~\ref{t:Delta-L}.
Soft potentials like the exponential one lead to smaller numbers.
Indeed, the variation $\partial \varphi / \partial \mu$ with the
chemical potential shows, on the one hand, an increased density 
in the bulk region. On the other hand, this is balanced by
the redistribution of the condensate density towards the tunnelling 
range $z < 0$, as can be seen in Fig.\:\ref{fig:condensates}(right).
\begin{table}[t]
\begin{center}
   \begin{tabular}{@{}l ll ll l} 
    \hline
	hard wall
	& \multicolumn{2}{c}{quarter-pipe}
	& \multicolumn{2}{c}{exponential}
	\\
    \hline
	$\Delta L = 1/(2\kappa)$ 
	& $0.63 / \kappa$	& $\kappa R = 3$
	& $0.046 / \kappa$	& $a = 1.4\,\kappa$
	\\
	& $0.72 / \kappa$	& $\kappa R = 5.5$
    & $0.028 / \kappa$	& $a = 0.7\,\kappa$ 
    \\
    & $0.78 / \kappa$ 	& $\kappa R = 8$ 
    & $0.042 / \kappa$	& $a = 0.4\,\kappa$
	\\	
    \hline
    \vspace*{-5ex}
   \end{tabular}
\end{center}
\caption[]{Missing length parameter $\Delta L$ in
$1/\mu'_0 = \partial N / \partial \mu = (L - \Delta L)/g$.
Exact result for the hard-wall potential.
In the exponential potential, $\Delta L$ goes through a minimum
for $a \approx 0.7\,\kappa$. 
The numerical calculations compute $\partial\varphi/\partial\mu$ by 
solving Eq.\,(\ref{eq:H3-phi-a-and-phi}) with a finite-difference
scheme [see Fig.\:\ref{fig:condensates}(right)].
The length $\Delta L$ is then proportional to the ``missing density'' 
in the space integral of $2g\varphi \partial\varphi/\partial\mu$,
similar to Eq.\,(\ref{eq:energy-0-extensive-plus-border}) and
Fig.\:\ref{fig:surface-tension-0}.
}
\label{t:Delta-L}
\end{table}

The physical content of these parameters becomes clear when $\hat{\psi}_c$
from Eq.\,(\ref{eq:psi-zero-mode}) is inserted into the energy 
operator~(\ref{eq:Omega-2nd-order}). 
The ground state contribution arises from the commutator between 
$\hat{Q}, \hat{P}$, as shown in the \reftoarxiv~A.2
\begin{equation}
H^{(2)}_c = 
\frac{\mu'_0}{2} \hat{P}^2 
- \frac{g}{2(L - \Delta L)} 
\int\limits^L\!{\rm d}z \, |\varphi|^2
\label{eq:ground-state-energy-2-c}
\end{equation}
The phase operator does not appear here, in agreement with the U(1) symmetry
of the problem \cite{Mora03}.
The first term vanishes in the ground state, and the second is
$- g N / 2L = -\mu/2$, up to corrections of relative order $\Delta L / L$
that vanish in the large-$L$ limit. 
This contribution to the ground state energy originates from the 
bulk. Mora and Castin have argued that it takes into account 
the actual number $N(N-1)/2$ of particle pairs in the condensate
and their interaction energy \cite{Mora03}, see also
\reftoarxiv~A.3.
It is curious to note that the surface 
energy~(\ref{eq:surface-energy-quarter-pipe}), being of opposite
sign, nearly compensates this contribution, although this 
occurs for a rather fictitious system with a single boundary.


\section{Conclusion}

We have analysed a fluctuation-induced surface energy in a one-dimensional 
system of ultracold Bosons
that does not reduce to the sum familiar from Casimir physics of
shifted zero-point energies per mode.
One rather needs the behaviour of the spatial mode functions near the
boundary.
Without restricting the dispersion relation of condensate fluctuations
to be linear,
we have combined numerical calculations and approximations 
based on a homogeneous system with locally the same density.
The surface energy in this system is dominated by the distortion
in the condensate wave function, while the contribution of virtual quasi-particles
amounts to about a fraction of one particle.
This is the level of accuracy that would be needed if a modelling
beyond mean-field theory is aimed at.
This has been done within the Yang-Yang thermodynamics 
of the Lieb-Liniger model \cite{Reichert_2019a, Reichert_2019b, YangYang69}, 
but restricted so far to a Dirichlet boundary condition.
It would be interesting to exploit the temperature-dependent excitation spectrum 
of this model and to extend it to a trap geometry with a soft confinement.
%

\subsection{Acknowledgments} 
I thank Anja Seegebrecht, Enrico Reiß, and Zoran Ristivojevic for inspiring discussions. 
This research was funded by the Deutsche Forschungsgemeinschaft 
(German Research Foundation) within SFB 1636, ID 510943930,
Projects No.\ A01 and A04.

%
\section{Appendix}
\renewcommand{\theequation}{\mbox{A.\arabic{equation}}}
\setcounter{equation}{0}

\subsection{\mbox{\bf A.1} Diagonalisation of fluctuation Hamiltonian}
\label{a:diagonalise-H2}

We check here that the expansion~(\ref{eq:Ansatz-for-psi})
of the operator $\hat{\psi}$ diagonalises the fluctuation Hamiltonian
$H^{(2)}$, the spatial integral of Eq.\,(\ref{eq:Omega-2nd-order}).
Exploiting the U(1) invariance of the theory, we may assume $\varphi$
and all the
$u_\omega$, $v_\omega$ to be real.
We insert the integral over Bogoliubov modes and find the following sum
of terms that multiply $\hat{\psi}^\dag$ from the right
\begin{eqnarray}
&& \displaystyle - \frac{\hbar^2}{4m} 
\frac{ {\rm d}^2 }{ {\rm d}z^2 } \left( u b + v b^\dag \right)
+ \frac{1}{2} (V - \mu) \left( u b + v b^\dag \right)
\nonumber\\
&& \displaystyle 
{} + \frac{g}{2} \varphi^2 \left( u b^\dag + v b \right)
\label{eq:expand-H2-one-half}
\end{eqnarray}
to be integrated over $z$ and $\omega$ (the frequency parameter 
is not shown explicitly). Collecting the coefficients of
$b$ and $b^\dag$, we recognise the first and second line of the eigenvalue problem~(\ref{eq:BdG}), so that this expression becomes
$\frac12\hbar\omega \left(u b - v b^\dag \right)$.
This yields the first line of Eq.\,(\ref{eq:expand-H2-step-1}).
The remaining terms can be written as the Hermitean conjugate of $\hat{\psi}^\dag$
times expression\,(\ref{eq:expand-H2-one-half}) and reduce to the second line
of Eq.\,(\ref{eq:expand-H2-step-1}).

When the spatial integration is performed in $H^{(2)}$, one uses
the orthogonality relations~(\ref{eq:uv-orthogonal}) to produce a
$\delta(\omega - \omega')$ that removes one frequency integral.
The same appears from the commutator arising from putting the product 
of hole operators into normal order,
\begin{equation}
- v_{\omega'} v_{\omega} \big[ b^{}_{\omega'}, \, b^\dag_{\omega} \big] 
= - |v_{\omega}|^2 \delta( \omega - \omega')
\label{eq:}
\end{equation}
but here, the spatial integral remains to be done. These manipulations produce
Eqs.\,(\ref{eq:expand-H2-step-1}, \ref{eq:ground-state-energy-0+2}) for
$H^{(2)}$.
The operator products $b b'$ and $b^\dag b^{\prime\dag}$ are found to be proportional 
to
\begin{equation}
\int\!{\rm d}z \left( u_\omega v_{\omega'} - v_\omega u_{\omega'} \right) = 0
\label{eq:uv-orthogonal-2}
\end{equation}
This identity follows from the observation that the mode pair
$(v, u)$ with exchanged particle and hole amplitudes solves
the Bogoliubov equations~(\ref{eq:BdG}) with the eigenvalue $-\omega$
when the orthogonality relation~(\ref{eq:uv-orthogonal}) is extended
to negative frequencies \cite{PitaevskiiStringari, Mora03}.

\subsection{\mbox{\bf A.2} Condensate phase and number fluctuations}
\label{a:psi-c-fluctuations}

When the fluctuation operator $\hat{\psi}_c$ [Eq.\,(\ref{eq:psi-zero-mode})]
for the condensate mode is inserted into the 
Hamiltonian~(\ref{eq:Omega-2nd-order}), we get the expression
($\hbar^2 / 2m = 1$ for simplicity)
\begin{eqnarray}
&& -\frac{{\rm i}\hat{Q}}{2} \Big[
	{-} \frac{ {\rm d}^2 \varphi }{ {\rm d}z^2 } 
	+ (V_2 - \mu) \, \varphi
	- g \varphi^3
	\Big]
\nonumber\\
&& {}
+ \frac{\hat{P}}{2} \Big[
	{-} \frac{ {\rm d}^2 \varphi_a }{ {\rm d}z^2 } 
	+ (V_2 - \mu) \, \varphi_a
	+ g \varphi^2 \varphi_a
	\Big]
\label{eq:expand-H2-2}
\end{eqnarray}
multiplying $\hat{\psi}_c^\dag$ from the right. The first line is
recognised as the Gross-Pitaevskii equation (GPE) for $\varphi$ and vanishes.
The second line can be written as the $\mu$-derivative of the GPE,
\begin{equation}
	- \frac{ {\rm d}^2 }{ {\rm d}z^2 } \frac{\partial\varphi}{\partial\mu}
	+ (V - \mu) \, \frac{\partial\varphi}{\partial\mu}
	+ 3 g \varphi^2 \frac{\partial\varphi}{\partial\mu}
	= 
	\varphi
\label{eq:H3-phi-a-and-phi}
\end{equation}
recalling the adjoint mode $\varphi_a$ to be proportional to $\partial\varphi/\partial\mu$
[Eq.\,(\ref{eq:phi-a-mu-derivative})]. The second square bracket in 
Eq.\,(\ref{eq:expand-H2-2}) then simply becomes $\mu'_0 \varphi$.
The remaining terms in the energy density are Hermitean conjugate to this one,
and we get
\begin{eqnarray}
{\cal H}^{(2)}_c 
&=& \frac{\mu'_0}{2} \left( \hat{\psi}_c^\dag \hat{P} \varphi + \hat{P} \varphi \hat{\psi}_c \right)
\nonumber\\
&=& \frac{\mu'_0}{2} \left( 2 \hat{P}^2 \varphi_a \varphi 
+ {\rm i} \big[ \hat{Q}, \hat{P} \big] \varphi^2 \right)
\label{eq:}
\end{eqnarray}
The space integral of the first term gives the unit overlap 
of Eq.\,(\ref{eq:phi-a-projector}).
Adding the commutator from the second term yields 
Eq.\,(\ref{eq:ground-state-energy-2-c}). 

\subsection{\mbox{\bf A.3} Correcting the number of particle pairs}
\label{a:N(N-1)}

The mean-field potential $g \varphi^2$ in the GPE~(\ref{eq:GPe}) makes a small
error because in pairwise interactions, a given particle only interacts with
$N-1$ others. This argument provides an alternative motivation for the 
fluctuation energy in Eq.\,(\ref{eq:ground-state-energy-2-c}) due to the 
condensate operator. 
We correct the GPE
\begin{equation}
- 
 \frac{ {\rm d}^2\tilde{\varphi} }{ {\rm d}z^2 }
+ V(z) \, \tilde{\varphi}
+ g (1-1/N)|\tilde{\varphi}(z)|^2 \, \tilde{\varphi}
= \mu \, \tilde{\varphi}
\label{eq:GPe-1/N}
\end{equation}
and take $1/N$ as a small parameter. The solution
$\tilde{\varphi}$ can be found by expanding around $\varphi$ which solves
Eq.\,(\ref{eq:GPe-1/N}) with $g$ in place of $g(1-1/N)$.
In terms of the correction $\tilde{\varphi}^{(2)} = \tilde{\varphi} - \varphi$,
\begin{equation}
\Big[
{-} 
 \frac{ {\rm d}^2 }{ {\rm d}z^2 }
+ V - \mu
+ 3 g |\varphi|^2 
\Big] \, {\varphi}^{(2)}
= (g/N) |\varphi|^2 \varphi
\label{eq:}
\end{equation}
In the bulk limit where $\tilde{\varphi}$ is constant, the derivatives
may be neglected. This motivates the so-called Thomas-Fermi solution with 
$V - \mu + 3 g|\varphi|^2 \approx 2 g |\varphi|^2$, and we find
$\tilde{\varphi}^{(2)} \approx \varphi / (2N)$.
The reduction of the interaction energy hence 
allows for a slightly larger particle density.
With this model, the ground state energy density becomes
\begin{eqnarray}
- \frac{g(1 - 1/N) }{2} |\tilde{\varphi}|^4 
&\approx& 
- \frac{g}{2} |\varphi|^4
- \frac{g}{2N} |\varphi|^4 
\label{eq:}
\end{eqnarray}
where the corrections in prefactor and condensate mode partially cancel.
With $|\varphi|^2 \approx N/L$, this gives 
the same scaling as in Eq.\,(\ref{eq:ground-state-energy-2-c}).




\end{document}